# Micrometer-Scale Ballistic Transport in Encapsulated Graphene at Room Temperature

A. S. Mayorov[1*], R. V. Gorbachev[1], S. V. Morozov[1,2], L. Britnell[1], R. Jalil[3],
L. A. Ponomarenko[1], P. Blake[3], K. S. Novoselov[1], K. Watanabe[4], T. Taniguchi[4], A. K. Geim[1,3]

[1]School of Physics and Astronomy, University of Manchester, Oxford Road, Manchester M13 9PL, UK
[2]Institute for Microelectronics Technology, 142432 Chernogolovka, Russia
[3]Manchester Centre for Mesoscience and Nanotechnology, University of Manchester, Manchester M13 9PL, UK
[4]National Institute for Materials Science, 1-1 Namiki, Tsukuba, 305-0044 Japan

*Devices made from graphene encapsulated in hexagonal boron-nitride exhibit pronounced negative bend resistance and an anomalous Hall effect, which are a direct consequence of room-temperature ballistic transport on a micrometer scale for a wide range of carrier concentrations. The encapsulation makes graphene practically insusceptible to the ambient atmosphere and, simultaneously, allows the use of boron nitride as an ultrathin top gate dielectric.*

In search for new phenomena and applications, which are expected, predicted or to be uncovered in graphene, it is important to continue improving its electronic quality that is commonly characterized by charge carrier mobility μ. Graphene obtained by mechanical cleavage on top of an oxidized Si wafer usually exhibits μ ~10,000 cm$^2$V$^{-1}$s$^{-1}$ [1]. For typical carrier concentrations $n \approx 10^{12}$ cm$^{-2}$, such quality translates into the mean free path $l = (h/2e)\mu(n/\pi)^{0.5}$ of the order of 100 nm where $h$ is Planck's constant and $e$ the electron charge. On the other hand, it has been shown that, if extrinsic scattering in graphene is eliminated, its mobility at room temperature ($T$) can reach ~200,000 cm$^2$V$^{-1}$s$^{-1}$ due to weak electron-phonon interaction [2]. Indeed, for $n$ ~10$^{11}$cm$^{-2}$, μ exceeding 100,000 cm$^2$V$^{-1}$s$^{-1}$ and 1,000,000 cm$^2$V$^{-1}$s$^{-1}$ at room and liquid-helium $T$, respectively, were demonstrated for suspended graphene annealed by high electric current [3-5]. However, suspended devices are extremely fragile, susceptible to the ambient atmosphere and difficult to anneal in the proper four-probe geometry (the latter was not achieved so far). Furthermore, it requires a significant amount of strain to suppress flexural modes in suspended graphene and retain high μ up to room $T$ [5]. Most recently, a breakthrough was achieved by using hexagonal boron-nitride (hBN) as an atomically smooth and inert substrate for cleaved graphene [6]. Such structures were shown to exhibit μ ~100,000cm$^2$V$^{-1}$s$^{-1}$ at $n$ ~10$^{11}$ cm$^{-2}$. Although μ achieved in graphene yield $l$ approaching 1 μm, no ballistic effects on this scale have so far been reported.

In this Letter, we describe devices made from graphene sandwiched between two hBN crystals. The devices exhibit room-$T$ ballistic transport over a 1 μm distance, as evidenced directly from the negative transfer resistance measured in the bend geometry [7]. At low $n$ ~10$^{11}$ cm$^{-2}$, the devices exhibit mobility μ >100,000 cm$^2$V$^{-1}$s$^{-1}$ even at room $T$, as determined from their response to gate voltage [1-6]. Moreover, $l$ continues growing with increasing $n$ and, at higher $n \approx 10^{12}$cm$^{-2}$, we find that our devices' longitudinal conductivity σ becomes limited by their width $w \approx 1$ μm rather than scattering in the bulk. From measurements of bend resistance $R_B$, we estimate that encapsulated graphene can exhibit μ ~500,000cm$^2$V$^{-1}$s$^{-1}$ and $l \approx 3$μm (at $n \approx 10^{12}$cm$^{-2}$) at low $T$, which rivals in quality the best suspended devices. In addition, the encapsulation has made graphene insusceptible to the environment and allows the use of hBN as an ultra-thin top gate dielectric.

The studied samples that we further refer to as graphene-boron-nitride (GBN) heterostructures were fabricated by using the following multistep technology. First, relatively thick (~10nm) hBN crystals were deposited on top of an oxidized Si wafer (100 nm of SiO$_2$). Then, sub-mm graphene crystallites were produced by cleavage on another substrate and transferred on top of the chosen hBN crystal by using alignment procedures similar to those described in ref. [6]. Electron-beam lithography and oxygen plasma etching were employed to define graphene Hall bars (see images in Figs. 1,2). The deposition of graphene on hBN resulted in numerous 'bubbles' containing trapped adsorbates (presumably hydrocarbons) and, if present in the active part of our devices, such bubbles caused significant charge inhomogeneity. This limited the achievable $w$ to ~1 μm, as we tried to fit the central wire inside areas free from the bubbles. The second hBN crystal (~10 nm thick) was carefully aligned to encapsulate the graphene Hall bar but leave the contact regions open for depositing metal (Au/Ti) contacts. In some devices, the top hBN crystal was used as a dielectric for local gating. After each transfer step, the devices were annealed at 300°C in an argon-hydrogen atmosphere to remove polymer residues and other contamination.

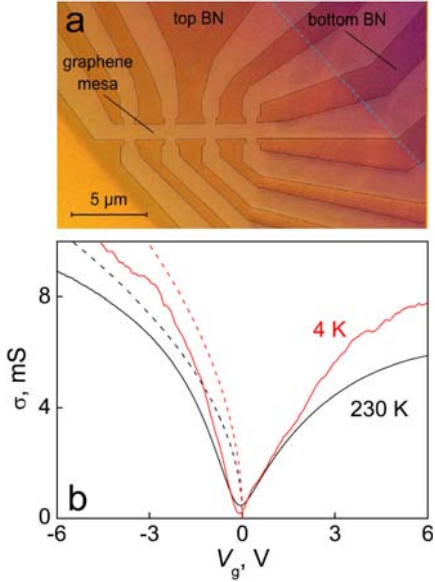

Figure 1: (**a**) – Optical micrograph of one of our GBN devices. The plasma etching resulted in a few-nm tall hBN mesa that could be visualized by using the differential interference contrast. To improve the mesa's visibility, its contour is shown by the thin grey lines. The slanted dashed line indicates the edge of the top hBN crystal. (**b**) – $\sigma(V_g)$ measured at two $T$ (solid curves). The dashed curves are $\sigma$ calculated by using the Landauer-Buttiker formula and numerical modeling of the transmission probability through a quantum wire with $w =1\mu$m. In the calculations, we assume diffusive boundary scattering and the internal mean free path in the graphene bulk $l_i$ =1.5 and 3 $\mu$m at 230 and 4 K, respectively, which are the values inferred from measurements of $R_B$ as described below.

Figure 1b shows $\sigma$ as a function of back-gate voltage $V_g$ for a GBN device, measured in the standard four-probe geometry. The minimum in $\sigma$ occurs at $V_g \approx$-0.1V, indicating little extrinsic doping (~$10^{10}$cm$^{-2}$). At small hole concentrations $n$ ~$10^{11}$cm$^{-2}$, the slopes of $\sigma(V_g)$ yield $\mu$ ≈140,000 and 100,000cm$^2$V$^{-1}$s$^{-1}$ at 4 K and room $T$, respectively (low-$n$ $\mu$ is about 30% lower for electrons). The values are in agreement with the measured Hall mobility. In general, at low $n$ our GBN devices exhibited $\mu$ between 20,000 and 150,000cm$^2$V$^{-1}$s$^{-1}$, tending to ≈100,000cm$^2$V$^{-1}$s$^{-1}$ in most cases. Another notable feature of Fig. 1b is a relatively weak $T$ dependence of $\sigma(V_g)$, which is surprising because electron-phonon scattering is expected to start playing a significant role in graphene of such quality [2,5,8]. Also, the strong sublinear behavior of $\sigma(V_g)$ is unusual for graphene measured in the four probe geometry. As shown below, these features are related to electron transport limited by boundary scattering so that $\sigma =2e^2/h(k_F l) \propto n^{1/2} \propto V_g^{1/2}$, where $l$ ~$w$ and, therefore, weakly depends on $T$. The importance of boundary scattering can also be appreciated if we estimate transmission probability $Tr$ through our devices (~3 $\mu$m long). To this end, the standard Landauer-Buttiker formula for quantum conductance $G =(4e^2/\pi h)(k_F w)Tr$ yields $Tr$ ≈0.3-0.5 at high $n$, which indicates quasi-ballistic transport.

To gain further information about electronic quality of the GBN bulk, we have studied bend resistance $R_B$ [9]. To this end, we applied current $I_{21}$ between contacts 2 and 1 and measured voltage $V_{34}$ between probes 3 and 4 (see Fig. 2), which yielded $R_B =R_{34,21}=V_{34}/I_{21}$. Different bend configurations (e.g., $R_{14,23}$ and $R_{32,14}$) yielded similar $R_B(V_g)$. For a diffusive conductor, $R_B$ should be equal to $\ln2/\pi\sigma$ [10]. The van der Pauw formula uses the diffusive approximation and can accurately describe $R_B(V_g)$ in the standard-quality graphene [9]. However, the formalism completely fails in our high-$\mu$ devices. Indeed, $R_B$ becomes negative which shows that most of the charge carriers injected from, say, contact 2 can reach contact 4 without being scattered. The counterintuitive negative resistance was observed in high-$\mu$ two-dimensional gases based on GaAlAs heterostructures and required $l_i$>>$w$ where $l_i$ is the mean free path in the bulk [7]. Such ballistic propagation of charge carriers has not been reported in graphene before, presumably due to its previously-insufficient quality (except for ref. [11] where low-$T$ nonlinear I-V characteristics were measured in 50-nm Hall crosses at a fixed $n$ and interpreted in terms of ballistic transport).

In contrast to the measurements in the standard geometry as in Fig. 1b, $R_B$ in Fig. 2a exhibits very strong $T$ dependence, in agreement with the expectations for high-$\mu$ graphene [5,8]. Despite this extra phonon scattering, $R_B$ remains negative at high $n$ for all $T$ ≤250 K and does not approach the gate dependence expected in the diffusive regime (dashed curve in Fig. 2a). This observation yields $l_i$ >$w$ ≈1 $\mu$m at room $T$, the condition essential for the observation of negative $R_B$ [7,12]. The strong $T$ dependence of $R_B$ also signifies that $l_i$ grows substantially with decreasing $T$. Complementary evidence for ballistic transfer through the Hall cross comes from devices with an extra barrier placed across one of the potential leads (Fig. 2b). When a voltage was applied to the narrow top gate, the potential barrier reflected carriers back into the cross and, accordingly, suppressed negative $R_B$. Also, note that, in the low-$n$ regime ($|V_g|$<0.5V) where we could determine $\mu$ from the linear dependence $\sigma \propto V_g$ as ~140,000cm$^2$V$^{-1}$s$^{-1}$, $R_B$ remains positive, as expected, because the corresponding $l = (h/2e)\mu(n/\pi)^{0.5} \leq 0.5\mu$m is insufficient for causing negative $R_B$.

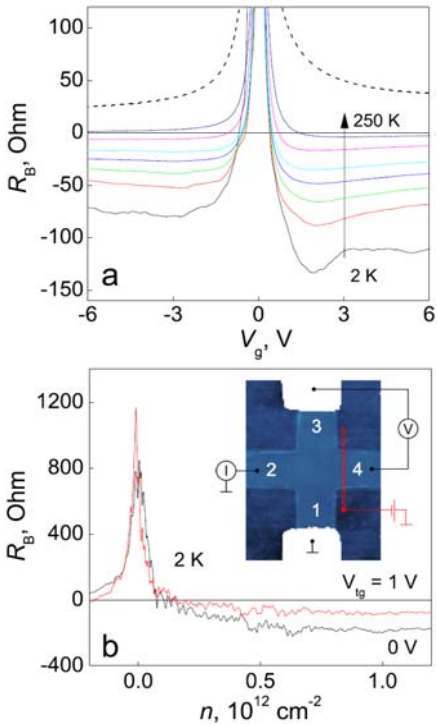

Figure 2: (**a**) – Bend resistance at various $T$ for the same device as in Fig. 1b. The curves from bottom to top correspond to 2, 50, 80, 110, 140, 200 and 250 K, respectively. The dashed curve is $R_B$ calculated using $\sigma(V_g)$ and the van der Pauw formula. (**b**) – Inset: atomic force micrograph of one of our Hall crosses. The scale is given by the device width $w \approx 1\mu m$. The drawings schematically depict the bend measurement geometry and a narrow top gate (in red) deposited across one of the leads at a later microfabrication stage. Main panel: $R_B(n)$ for a device with the such a top gate. The negative $R_B$ can be suppressed by applying top-gate voltage $V_{tg}$ which creates an extra barrier and reflects electrons.

To elucidate the micron-scale ballistic transport in our GBN heterostructures, Fig. 3a shows $R_B$ as a function of magnetic field $B$ applied perpendicular to graphene at a fixed $V_g$ (+3V in this case). As expected [7], $R_B$ changes its sign with increasing $B$ because injected electrons are bended by $B$ and can no longer reach the opposite contact ballistically. This behavior is in agreement with the one reported in GaAlAs heterostructures [7,13]. The characteristic field $B_0$ in Fig. 3a is ~0.1T, which corresponds to a cyclotron orbit of radius $r_c = \hbar(\pi n)^{1/2}/eB \approx 1\mu m$, that is equal to $w$, in agreement with theory [7,13] ($n \approx 6\times10^{11}$ cm$^{-2}$ in this case). Furthermore, ballistic transport is expected to cause an anomalous behavior of Hall resistivity $R_H$ such that it is no longer a linear function of $B$. Fig. 3b shows that, indeed, our devices exhibit nonlinear $R_H(B)$ with a notable kink at the same characteristic $B_0$. This anomaly is usually referred to as the last plateau and absent in diffusive systems. The kink almost disappears at room $T$ (Fig. 3b) indicating that we get closer to the diffusive regime. The functional form of $R_H(B)$ strongly depends on the exact shape of Hall crosses, and the anomaly becomes minor if a cross has sharp corners [7,13], as is the case of our devices (see image in Fig. 2b).

Figure 3: Ballistic transport in magnetic field. (**a**) – $R_B(B)$ for a fixed $n \approx 6\times10^{11}$ cm$^{-2}$. $T$ is 50, 80, 110, 140, 200 and 250 K (from bottom to top curves, respectively). Inset: $R_B(B)$ calculated for a Hall cross using the billiard-ball model [7] and scaled for the case of our graphene devices and the above $n$. (**b**) – Hall resistance $R_H$ measured at 50 and 250 K. Inset: $R_H(B)$ found theoretically for rounded corners [7] and scaled for our case. The red line in the inset indicates the diffusive limit.

The negative $R_B$, its magnetic field behavior, anomalies in $R_H$ and the influence of the top gate unambiguously prove that, in our Hall crosses, charge carriers can reach the opposite lead ballistically, without scattering. This yields $l$ longer than 1 μm for all $|V_g| >1V$ where the large negative $R_B$ is observed ($|n|\geq 2\times10^{11}$ cm$^{-2}$). To appreciate so large values of $l$, let us mention that in suspended devices [3,4] and graphene on BN [6], ultra-high μ were reported only at low $n \sim 10^{11}$ cm$^{-2}$ which translates into submicron $l$ [3,4], and $l \approx 1\mu m$ were achieved only in suspended devices with a million μ at low $T$ [5].

For $l_i >w$, the boundary scattering makes σ only weakly dependent on the bulk quality of graphene and, to obtain a better estimate for $l_i$ than just $\geq 1\mu m$ as above, we used numerical simulations. We calculated $R_B$ by using the billiard-ball model [7] and assuming diffusive boundary scattering. If the scattering is assumed specular, calculated $R_B$ cannot reach the large negative values observed experimentally. This agrees with general expectations that etched graphene edges are generally rough and scatter diffusively. Diffusive boundary scattering decreases $\sigma$ of a ballistic wire (transmission probability decreases) but makes $R_B$ more negative due to collimation effects [14]. This is consistent with our experiment that shows higher (more ballistic) $\sigma$ for holes but more negative $R_B$ for electrons and vice versa (cf. Figs. 1 and 2). This asymmetry can be attributed to a larger degree of diffusivity in boundary scattering for electrons, which implies an extra charge that breaks the electron-hole symmetry of the boundary. Under the assumption of diffusive scattering, the measured $R_B$ yield $l_i \approx 1.5 \mu m$ at room $T$ and $\approx 3 \mu m$ below 50K for $|n| > 2 \times 10^{11} cm^{-2}$ ($|V_g| > 1V$). Although the exact values are inferred by using the numerical modeling and assuming diffusive boundaries, such large $l_i$ are essential to explain qualitatively both large negative $R_B$ and its strong $T$ dependence (for example, $l_i \leq 1$ μm would be inconsistent with these observations). The inferred $l_i$ also allow us to understand the behavior of $\sigma$ and its weak $T$ dependence, and the dashed curves in Fig. 1b show $\sigma(V_g)$ calculated within the same model and parameters.

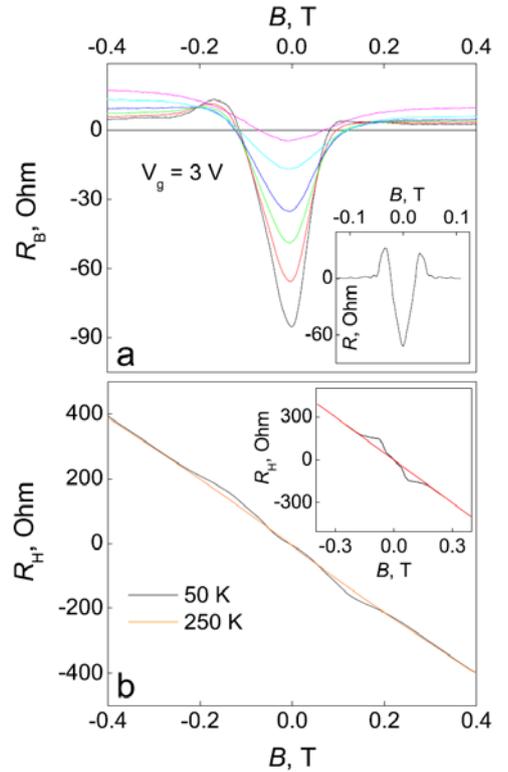

Finally, we note that, for $n \approx 4 \times 10^{11} cm^{-2}$ where $R_B$ reaches its most negative value, $l_i \approx 3 \mu m$ implies intrinsic $\mu$ ~500,000 $cm^2V^{-1}s^{-1}$. This is consistent with $\mu$ ~150,000 $cm^2V^{-1}s^{-1}$ found from the field effect at significantly lower $n \approx 1 \times 10^{11} cm^{-2}$ where charge inhomogeneity remains significant. The latter regime corresponds to $l_i \leq 0.5 \mu m$ and does not allow negative $R_B$, in agreement in the experiment. To confirm the above $\mu$ at high $n$ by using standard field-effect measurements would require GBN devices with $w > 5 \mu m$, which we have so far been unable to achieve because of the mentioned bubbles that result in charge inhomogeneity.

In conclusion, graphene encapsulated in hBN exhibits robust ballistic transport with a large negative transfer resistance and the mean free path exceeding ~3 μm at low $T$. Away from the neutrality point, (for carrier concentrations above $10^{11}$ $cm^{-2}$) the longitudinal conductivity of our 1 μm wide devices becomes limited by diffusive scattering at the sample boundaries. The demonstrated graphene-boron-nitride heterostructures is a further improvement with respect to the devices reported previously and shows the way to achieve million mobilities for graphene on a substrate.

Acknowledgement. This work was supported by the Körber Foundation, Engineering and Physical Sciences Research Council (UK), the Office of Naval Research, the Air Force Office of Scientific Research and the Royal Society.

* mayorov@gmail.com